\documentclass[proceedings,article,submit,moreauthors,pdftex,10pt,a4paper]{mdpi}

\preto{\abstractkeywords}{\nolinenumbers}
\firstpage{1} 
\pubvolume{xx}
\issuenum{1}
\articlenumber{1}
\pubyear{2018}
\copyrightyear{2018}
\history{Received: date; Accepted: date; Published: date}

\pdfoutput=1

\usepackage[english]{babel}
\usepackage[utf8]{inputenc}
\usepackage{enumerate}
\usepackage{braket}
\usepackage{amsfonts}

\Title{TWO-QUBITS IN A LARGE-S ENVIRONMENT}

\Author{Eliana Fiorelli $^{1,2,\dagger}$\orcidA{}, Alessandro Cuccoli $^{3,4}$ and Paola Verrucchi $^{5,3,4}$*}

\AuthorNames{Firstname Lastname, Firstname Lastname and Firstname Lastname}

\address{%
$^{1}$ \quad School of Physics and Astronomy, University of Nottingham, Nottingham, NG7 2RD, UK; eliana.fiorelli@nottingham.ac.uk\\
$^{2}$ \quad Centre for the Mathematics and Theoretical Physics of Quantum Non-equilibrium Systems, University of Nottingham, Nottingham NG7 2RD, UK;\\
$^{3}$ \quad Dipartimento di Fisica e Astronomia, Universit\`a di Firenze, I-50019, Sesto Fiorentino (FI), Italy; cuccoli@fi.infn.it\\
$^{4}$ \quad INFN, Sezione di Firenze, I-50019, Sesto Fiorentino (FI), Italy;\\
$^{5}$ \quad ISC-CNR, at Dipartimento di Fisica, Universit\`a di Firenze, I-50019, Sesto Fiorentino (FI), Italy; verrucchi@fi.infn.it}

\abstract {
We analytically express the loss of entanglement between the components 
of a quantum device due to the generation of quantum correlations with 
its environment, and show that such loss diminishes when the latter
is macroscopic and displays a semi-classical behaviour.
We model the problem as a device made of a couple of qubits
with a magnetic environment: this choice allows us to implement the above 
condition of semi-classical macroscopicity in terms of a large-$S$ 
condition, according to the well known equivalence between classical 
and $S\to\infty$ limit. A possible strategy for protecting internal 
entanglement exploiting the mechanism of domain-formation typical of 
critical dynamics is also suggested.
}

\keyword{entanglement, quantum devices, environment}
\begin{document}
\section{Introduction}

One of the most relevant issues in the design of quantum devices is to 
understand how their internal quantum components interface with the 
external macroscopic apparatus that allows us to control, and 
ultimately make use of, these extraordinary tools.
In fact, we know that entanglement between components (hereafter dubbed 
"internal entanglement") is key to the effective functioning of quantum 
devices. On the other hand, the above interface is expected to imply the 
dynamical generation of entanglement between the device and its 
environment (hereafter dubbed "external entanglement"), be it a control 
system, a measuring apparatus, or just a noisy surrounding.
Therefore, understanding how internal and external entanglement depend 
on each other, and whether it is possible to protect the former by 
reducing the latter, is important.
In this work we consider a quantum device $D$ and the apparatus $M$ that 
works as interface with the user. Although $M$ might in principle 
include other physical systems, such as a thermal bath, in this work we 
will not take this possibility into account. In fact, in order to obtain 
an analytical description of how entanglement internal to $D$ is 
affected by the presence of $M$, we further restrict ourselves to the 
case when $D$ is a qubit-pair and $M$ can be described in terms of two 
quantum objects with spin much larger than $1/2$ (so as to make them different from the qubits by definition).


\section{The model and its entanglement properties}

We study an isolated system 
$\Psi= D+M$, with $D = Q_{1}+Q_{2}$ and $M=A+B$,
where $Q_{1,2}$ are qubits, i.e. quantum systems whose Hilbert spaces 
are two-dimensional, while $A,B$ are such that 
dim${\cal H}_{\rm A,B}=2S_{\rm A,B}+1$, with $2S_{\rm A,B}$ 
integer numbers much larger than unity.
Overall it is 
dim${\cal H}_{D}=4$, 
dim${\cal H}_M=(2S_{\rm A}+1)(2S_{\rm B}+1)$, and 
dim${\cal H}_\Psi=4(2S_{\rm A}+1)(2S_{\rm B}+1)$.
In what follows the qubits will be described by the Pauli operators
$\hat{\boldsymbol{\sigma}}_i$ such that
$\left[\hat{\sigma}_i^{\lambda},\hat{\sigma}_j^{\mu} \right] =
i2\epsilon_{\lambda\mu\nu}\hat{\sigma}_i^{\nu}\delta_{ij}$, 
with $\lambda(\mu,\nu)=x,y,z$, and $i(j)=1,2$.
As for $A$ and $B$, they will be represented in terms of spin 
operators $\hat{\mathbf S}_*$ such that
$\left[\hat S^{\lambda}_*,\hat S^{\mu}_\#\right] =
i\epsilon_{\lambda\mu\nu}\hat S^{\nu}_*\delta_{*\#}$, with
$|\hat{\mathbf S}_*|^2=S_*(S_*+1)$, and $*(\#)=A,B$. 
The total system $\Psi$ is assumed isolated, and hence in a pure state 
$\ket{\Psi}$ at all time.
The device is prepared in a state featuring some internal entanglement, 
meaning that $Q_1$ and $Q_2$ are entangled. 
On the other hand, we take $S_A$ and $S_B$ independent from each 
other, and separately coupled with $D$, i.e. with $Q_1$ and $Q_2$, via 
an interaction upon which we do not make assumptions. 

In this setting, the state $\ket{\Psi}$ at whatever time after the 
interaction starts can be written as
\begin{equation}
\ket{\Psi}=\sum_{d=1}^4
\sum_{\alpha=1}^{2S_A+1}\sum_{\beta=1}^{2S_B+1}
g_{d\alpha}l_{d\beta}\Ket{d}
\otimes\Ket{\alpha}\otimes\Ket{\beta},
\label{e.PsiQ1Q2SASB}
\end{equation}
where  $\lbrace\Ket{d}\rbrace_{\mathcal{H}_D}$, 
$\lbrace\Ket{\alpha}\rbrace_{\mathcal{H}_A}$ and 
$\lbrace\Ket{\beta}\rbrace_{\mathcal{H}_B}$ are orthonormal bases for 
the Hilbert spaces of $D$, $A$, and $B$, respectively. The complex 
coefficients $\{g_{d\alpha}\}$ and $\{l_{d\beta}\}$ satisfy 
$\sum_{d\alpha\beta}\vert g_{d\alpha}l_{d\beta}\vert^{2} = 1$, due to 
the normalization of $\ket{\Psi}$. 
The form (\ref{e.PsiQ1Q2SASB}) ensures that the state of $M$ is 
separable \cite{LewensteinEtal08}, meaning that there is no
entanglement between $S_{1}$ and $S_{2}$. 
For the sake of a lighter notation and without loss of 
generality, we set
$S_A=S_B=S$.

Aim of the following analysis is to understand how the internal 
entanglement depends on the spin-values $S$, particularly 
when the condition of large-$S$ is enforced on both $A$ and $B$. 
To this respect, we remind that a spin-$S$ system exhibits 
a classical-like behaviour when $S\to\infty$\cite{Lieb71},
which implies the vanishing of any sort of quantum correlations, 
including entanglement, with other systems; we hence expect 
$M$ to disentangle from $D$ when $S$ increases.
In order to check whether this process can effectively protect the 
internal entanglement we proceed as follows.

We first conjecture that the coefficients entering the state 
\eqref{e.PsiQ1Q2SASB} have the following form in the large-$S$ limit
\begin{equation}
g_{d\alpha}l_{d\beta} =\frac{1}{N(S)} 
c_d(1+x_{d\alpha}(S))(1+y_{d\beta}(S))~, 
\label{e.coefSASB} 
\end{equation} 
where $x_{d\alpha}(S)$ and $y_{d\beta}(S)$ are 
decreasing functions of $S$, $\forall d,\alpha,\beta$, such 
that 
\begin{equation} 
\lim_{S\rightarrow\infty} x_{d\alpha}(S)= 
\lim_{S\rightarrow\infty} y_{d\beta}(S)=0~,~\forall d,\alpha,\beta~,  
\label{e.largeS}
\end{equation} 
while $N(S)$ 
is the normalization factor that goes to 1 in the large-$S$ limit.
Moreover, in order to make our analytical results more readable, we 
restrict ourselves to the case when the qubit pair is confined to 
the two-dimensional Hilbert subspace generated 
by two of the four states $\{\ket{d}\}$, namely
$\ket{d=3}=\ket{00}$ and $\ket{d=4}=\ket{11}$ (we choose the indexes 3 
and 4 to avoid confusion with the qubit labels, 1 and 2).

We can now determine the explicit expression for the concurrence $C_{Q_1Q_2}$
relative to the state $\rho_D= {\rm Tr}_M\ket{\Psi}\bra{\Psi}$, 
which is a proper measure\cite{Wootters98} of the internal 
entanglement, i.e. the one between $Q_1$ and $Q_2$.
Notice that the concurrence can be here used to study the internal 
entanglement because $D$ is made of a qubit-pair, i.e. the only system 
for which the concurrence relative to a mixed state is defined. 

Referring to the state \eqref{e.PsiQ1Q2SASB}, using the form 
\eqref{e.coefSASB} with $d=3,4$ only, and understanding the $S$ 
dependence, we finally find
\begin{equation}
C_{Q_{1}Q_{2}}=
\max \left\lbrace 0, \frac{2\vert c_{3}
c_{4}\vert}{N^{2}}\left|
\sum_{\alpha}(1+x_{3\alpha})(1+x^*_{4 \alpha}) 
\sum_{\beta}(1+y_{3\beta})(1+y^*_{4 \beta})\right|\right\rbrace~.
\label{e.CQ1Q2}
\end{equation}
Another useful entanglement measure that we take into consideration 
is the one-tangle $\tau_Q$, which quantifies the entanglement between a
qubit and whatever else determines its state $\rho_Q$, according to 
$\tau_Q=4{\rm det}\rho_Q$\cite{CoffmanKW99}. In our setting, we use
it to evaluate how much entanglement one of the two qubits of $D$
shares with the system made of $M$ and the other qubit. 
Its explicit form reads
\begin{equation}
\tau_{Q_{1}} = \
\frac{4\vert c_{3}c_{4}\vert^{2}} {N^{4}}
\sum_{\alpha}|1+x_{3\alpha}|^2 
\sum_{\alpha'}|1+x_{4\alpha'}|^2 
\sum_{\beta}|1+y_{3\beta}|^2
\sum_{\beta'}|1+y_{4\beta'}|^2=\tau_{Q_2}~,
\label{e.tauQ1}
\end{equation}
where the last equation follows from the symmetry of the 
setting w.r.t. to the swap $Q_1\leftrightarrow Q_2$.
It is important to notice that while the concurrence $C_{Q_1Q_2}$ 
quantifies just the 
useful internal entanglement that allows the device $D$ to function 
efficiently, the one tangle $\tau_{Q_1}$ incorporates some 
useless external entanglement, and comparing 
the twos can help quantifying the detrimental effect of $M$ upon the 
qubit-pair entanglement, as further commented upon in the concluding 
section.

In order to evaluate $C_{Q_1Q_2}$ and $\tau_{Q_1}$ from 
Eqs.\eqref{e.CQ1Q2} 
and \eqref{e.tauQ1} one has to choose 
the coefficients $\{x_{d\alpha}\}$, and $\{y_{d\beta}\}$.
In order to keep our analysis as general as possible, for each value of  
$S$, we have repeated the calculation of both quantities for 200 times, each 
time using a different set of coefficients, 
$\{x_{3\alpha},x_{4\alpha},y_{3\beta},y_{4\beta}\}$ randomly generated
according to $x_{i\alpha}\in(0,x_{\rm max}]$ and 
$y_{i,\beta}(0,y_{\rm max}]$ , $i=3,4$, $\forall \alpha,\beta$.
The average of the 200 values thus obtained for
$C_{Q_1Q_2}$, and $\tau_{Q_1}$, is then taken as the, respective, 
proposed result. In fact, reminding the $S$-dependence that we have 
understood in the above coefficients, we have further enforced the condition 
\eqref{e.largeS} by taking $x_{\rm max}=\frac{1}{2S^n}$, with $n=1,2,3$, 
and the same for $y_{\rm max}$. 
As for $c_3$ and $c_4$ we have put them both equal to $1/\sqrt{2}$.  

\begin{figure}[h!] \centering 
\includegraphics[width=10cm]{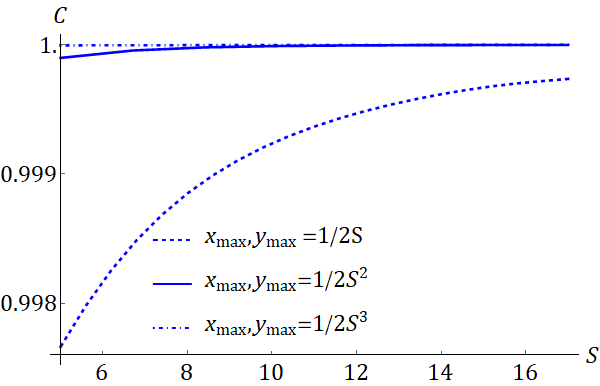}
\llap{\makebox[1.0cm][c]{\raisebox{0.75cm}
{\includegraphics[width=6cm]{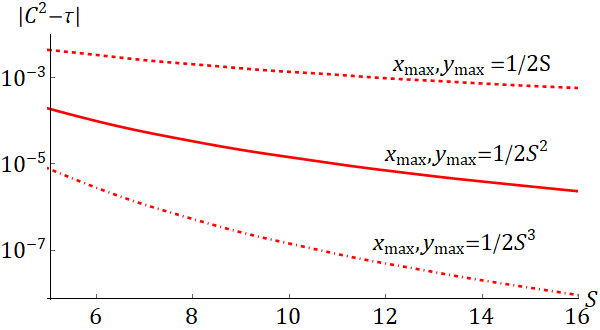}}}}
\caption{\small{ $C_{Q_{1}Q_{2}}$ as a function of $S$. In the inset 
$C_{Q_{1}Q_{2}}^{2}-\tau_{Q_{1}}$ as a function of $S$. Each line 
correspond to a specific choice of $x_{max}$ and 
$y_{max}$ (see text).}}
\label{f.fig1}
\end{figure}

In Fig.\ref{f.fig1} we show $C_{Q_1Q_2}$ as a function of $S$, and 
$n=1,2,3$. We see that, even in the worse case, $n=1$, the internal 
entanglement increases with $S$.
In order to check whether a larger internal entanglement can be actually 
due to a reduction of the 
external one, in the inset of Fig.\ref{f.fig1} we show the difference 
$|C^2_{Q_1Q_2}-\tau_{Q_1}|$, that 
provides an estimate of the internal-entanglement squandering due 
to the onset of quantum correlations between $D$ and $M$.
It is clearly seen, both for $n=1$ and 2, that a large value of $S$  
prevents the above onset, resulting in an effective protection of 
internal entanglement.
 
\section{Discussion}
In the above section we have introduced the idea that 
taking a large value of $S$ might help protecting the internal 
entanglement, as it induces a classical-like behaviour for $M$, and 
hence a net reduction of its quantum correlations with $D$.
However, this argument only works if one assumes that some constraint 
upon the entanglement between, say, $Q_1$, and other quantum systems 
holds. In fact, one such constraint exists and usually goes under the 
name of "monogamy of entanglement", analytically expressed by 
inequalities taking different forms depending on the specific case 
considered.
In the case of $N$ qubits in a pure state, it is expressed by
\begin{equation}
\sum_{i=2}^N C^2_{Q_1Q_i}\le \tau_{Q_1R}\le 1~,
\label{e.monoEnta}
\end{equation}
where $\tau_{Q_1R}$ is the one-tangle between $Q_1$ and the other $N-1$ 
qubits. Although the above expression does not fit our situation, as we 
are not dealing with $N$ qubit, we can use it as follows (we still take 
$S_A=S_B$ for the sake of simplicity). 

In order for the physical objects that model $M$ to be
described as spin-$S$ systems they must be made of a set 
of qubits $\{q^*_i\}$, with $i=1,...N\ge 2S$ and $*=A,B$, coupled 
amongst themselves in a way such that the total spin of the set keeps 
the constant value $S$.
In our setting, this translates Eq.\eqref{e.monoEnta} into 
\begin{equation}
C^2_{Q_1Q_2}
+\sum_{i=1}^N C^2_{Q_1q^A_i}+\sum_{i=1}^N C^2_{Q_1q^B_i}
\le\tau_{Q_1}\le 1~.
\label{e.monoNostra}
\end{equation}
Although we cannot limit the sums 
entering the above equation by using Eq.\eqref{e.monoEnta} again, as 
this exclusively hold for qubits in a pure state while 
$Q_1+\{q_i^A\}$ is in a mixed one,  yet we can understand that one 
possibility for Eq.\eqref{e.monoNostra} to stay meaningful as 
$S\to\infty$, i.e. $N\to\infty$, is that both sums it contains do 
vanish. In fact, this can be analytically demonstrated by
enforcing the condition 
\eqref{e.largeS} into Eqs.\eqref{e.CQ1Q2} and \eqref{e.tauQ1} via
neglecting all powers of order 2 and higher in the coefficients 
$\{x_{d\alpha}\}$ and $\{y_{d\beta}\}$.
This leads to
\begin{equation}
C_{Q_{1}Q_{2}}^{2}\sim 
\frac{4(2S+1)^2\vert c_{3}c_{4}\vert^{2}}{N}
\sum_{\alpha,\beta} \left[ 
1+2{\rm Re}\left( \bar{x}_{1\alpha } + \bar{x}_{4\alpha} + 
\bar{y}_{1\beta} + \bar{y}_{4\beta}\right) \right]\sim\tau_{Q_1}~,
\label{e.largeS_C2&tau}
\end{equation}
which implies that $Q_1$ is 
entangled almost exclusively with $Q_2$ if $S\to\infty$ as $M$ becomes 
macroscopic, consistently 
with the idea that large-$S$ systems do not share quantum correlations\citep{FotiCV16}.

This work confirms that a good strategy for protecting the internal 
entanglement of a quantum device $D$ is that of making its 
control/reading apparatus $M$ to feature a semi-classical behaviour, by 
this meaning that it still admits a quantum-mechanical description, so 
as to keep talking with the device, but with genuinely quantum 
properties, such as entanglement, already on the verge of depletion.
In fact, in order to test this strategy, we are specifically considering 
the case when $M$ is a many-body spin-system dynamically driven near a 
quantum critical point, where the mechanism of domain-formation might 
indeed induce a semi-classical behaviour as described above. Related 
work is in progress.

\acknowledgments{
This work is done in the framework of the Convenzione operativa
between the Institute for Complex Systems of the Italian
National Research Council (CNR), and the Physics and Astronomy 
Department of the University of Florence. We acknowledge financial support from the University of Florence in the framework of the University Strategic Project Program 2015 (project BRS00215).}
\reftitle{References}
\bibliographystyle{mdpi}
\bibliography{biblio}

\end{document}